\documentclass[conference,a4paper]{IEEEtran}

\usepackage{amsmath,epsfig,amssymb,verbatim}
\usepackage{cite}
\usepackage{graphicx}
\usepackage{bm,bbm}
\usepackage{xcolor}
\usepackage{subfigure}
\usepackage{stackengine}
\usepackage{cleveref}
\usepackage{multicol}
\usepackage{multirow}
\def\delequal{\mathrel{\ensurestackMath{\stackon[1pt]{=}{\scriptstyle\Delta}}}}

\newcommand{\E}{\mathbb{E}}

\newcommand{\Lp}{\mathcal{L}}

\newtheorem{Theorem}{Theorem}
\newtheorem{Lemma}{Lemma}

\def\BibTeX{{\rm B\kern-.05em{\sc i\kern-.025em b}\kern-.08em
   T\kern-.1667em\lower.7ex\hbox{E}\kern-.125emX}}

\begin{document}
\title{Impact of Pointing Error on Coverage Performance of 3D Indoor Terahertz Communication Systems}

\author{\IEEEauthorblockN{Zhifeng Tang$^{\ast}$, Nan Yang$^{\ast}$, Xiangyun Zhou$^{\ast}$, Salman Durrani$^{\ast}$, 
Markku Juntti$^{\dagger}$, and Josep Miquel Jornet$^{\ddagger}$}
\IEEEauthorblockA{$^{\ast}$School of Engineering, Australian National University, Canberra, ACT 2600, Australia}
\IEEEauthorblockA{$^{\dagger}$Centre for Wireless Communications, University
of Oulu, Oulu 90014, Finland}
\IEEEauthorblockA{$^{\ddagger}$Department of Electrical and Computer Engineering, Northeastern University, Boston, MA 02120, USA}
\IEEEauthorblockA{Emails: \{zhifeng.tang, nan.yang, xiangyun.zhou, salman.durrani\}@anu.edu.au, \\markku.juntti@oulu.fi, j.jornet@northeastern.edu}
\thanks{An extended version of this work has been submitted}\vspace{-3.2em}
}
\maketitle

\begin{abstract}
    In this paper, we develop a tractable analytical framework for a three-dimensional (3D) indoor terahertz (THz) communication system to theoretically assess the impact of the pointing error on its coverage performance. Specifically, we model the locations of access points (APs) using a Poisson point process, human blockages as random cylinder processes, and wall blockages through a Boolean straight line process. %To address the high penetration loss caused by blockages, we adopt the nearest line-of-sight AP association strategy.  
    A pointing error refers to beamforming gain and direction mismatch between the transmitter and receiver. We characterize it based on the inaccuracy of location estimate. We then analyze the impact of this pointing error on the received signal power and derive a tractable expression for the coverage probability, incorporating the multi-cluster fluctuating two-ray distribution to accurately model small-scale fading in THz communications. Aided by simulation results, we corroborate our analysis and demonstrate that the pointing error has a pronounced impact on the coverage probability. Specifically, we find that merely increasing the antenna array size is insufficient to improve the coverage probability and mitigate the detrimental impact of the pointing error, highlighting the necessity of advanced estimation techniques in THz communication systems.
\end{abstract}


\section{Introduction}

Terahertz (THz) communications has emerged as a promising solution for the sixth generation wireless systems to meet the growing demands for ultra-high-speed and high-capacity wireless connectivity \cite{6GNet2020}. Specifically, THz communications operate within the 0.1–10 THz frequency range, offering abundant spectrum resources and ultra-short symbol durations \cite{THzComMCS2024}. However, several technical challenges hinder its widespread adoption, including severe spreading loss, molecular absorption, and sensitivity to blockages \cite{tang2025lowcomplexityartificialnoise}. To unlock the potential of THz technology for next-generation wireless systems, it is essential to address the challenges through comprehensive system analysis, optimization based design \cite{Jornet2011}.

%%%%%%%%%%%%%%%%%%%%%%%%%%%%% 

Recent studies have evaluated the performance of THz communication systems while accounting for the unique characteristics of THz channels \cite{Venugopal2016Access,Petrov2017,Wu2021TWC,Shafie2021JSAC,Kouzayha2023twc,Tang2024Arxiv}. In particular, \cite{Venugopal2016Access} analyzed the impacts of human blockages on signal propagation and derived a closed-form expression for the coverage probability. In \cite{Petrov2017}, a Taylor series approximation was employed to examine the interference and coverage probability in THz communication systems. The impact of wall blockages on the coverage performance was examined in \cite{Wu2021TWC}, emphasizing their significance in indoor environments. Furthermore, \cite{Shafie2021JSAC} investigated the joint impact of human and wall blockages on the coverage performance of a three-dimensional (3D) THz communication system. Similarly, \cite{Kouzayha2023twc} assessed this joint impact on the coverage probability and average transmission rate in integrated sub-6 GHz and THz systems. More recently, \cite{Tang2024Arxiv}  incorporated a fluctuating two-ray (FTR) model to accurately capture small-scale fading characteristics and analyzed the impact of user location on coverage performance in indoor environments.
%examined the impact of user location on coverage performance in indoor environments, incorporating a fluctuating two-ray (FTR) model to more accurately capture the small-scale fading characteristics in THz communications. 

It is worth noting that most of these studies have relied on high-directional antennas to mitigate path loss and enhance signal power in THz communication systems. While such antennas improve directivity and beamforming capabilities, they are highly susceptible to transceiver misalignment, known as pointing error, which significantly degrades system performance \cite{Boulogeorgos2022,Petrov2020tvt}. Therefore, several studies have analyzed the impacts of the pointing error on system performance in various THz communications scenarios. For instance, \cite{Dabiri2022wcl} introduced an analytical framework to evaluate the pointing error in THz links, providing fundamental insights into the severity of the pointing error. Building on this, \cite{Safahan2025coml} proposed an analytical pointing error model for highly directional THz transmissions, facilitating performance evaluation in practical deployments. Additionally, \cite{Dabiri2023ojcom} examined the impact of the pointing error on the outage probability in an unmanned aerial vehicle (UAV)-assisted communications system and proposed optimization algorithms to ensure robust connectivity in the presence of mobility-induced misalignment. While these studies provide valuable insights into the pointing error in various cases, its impact on the coverage performance in indoor THz communication systems has not been adequately evaluated, which motivates this work.

In this paper, we introduce a tractable analytical framework to precisely assess the impact of the pointing error on the coverage probability in indoor THz communication systems. We model the locations of access points (APs) using a Poisson point process, human blockages as a random cylinder process, and wall blockages through a Boolean straight-line process. The pointing error is characterized by location estimation inaccuracies between the transmitter and receiver. Focusing on this system model, we first analyze the impact of the pointing error on the received signal and then employ stochastic geometry to derive a tractable expression for the coverage probability. Aided by simulations, we demonstrate the accuracy of our analytical results and reveal the substantial impact of pointing errors on coverage performance, highlighting that coverage probability is fundamentally limited by the pointing error and cannot be indefinitely improved by merely increasing the antenna array size.
%of our analytical results and quantify the significant impact of the pointing error on the coverage performance, which implies that the coverage probability is fundamentally constrained by the pointing error and cannot be indefinitely improved by simply increasing the antenna array size.

\section{System and Channel Model}\label{Sec:System}

\subsection{System Deployment}

We consider a generalized 3D indoor THz communication system, where multiple APs mounted on the ceiling transmit THz signals to user equipment (UEs). Specifically, we assume that the ceiling has a fixed height $h_A$ in the indoor environment and the locations of THz APs follow a Poisson point process (PPP) with the density of $\lambda_A$. We also assume that UEs, having a fixed height $h_U$, are randomly distributed on the ground. In this system, we randomly select one UE and refer to it as the typical UE, denoted by $U_0$.

%\begin{figure}[t]
%    \centering
%        \subfigure[The top view.]{
%            \includegraphics[width=0.42\columnwidth]{HumanBlockV.jpg}
%            }\label{fig:TopviewHB}
%        \subfigure[The vertical view.]{
%            \includegraphics[width=0.42\columnwidth]{HumanBlockH.jpg}
%            }\label{fig:VertviewHB}
%            \vspace{-0.5em}
%        \caption{Top and vertical views of human blockage for an AP-UE link.}\vspace{-1.5em}
 %       \label{fig:HB}
%\end{figure}

We consider that the blockage of a transmission link is caused by both human bodies and wall blockages. Following state-of-the-art studies \cite{Wu2021TWC,Tang2024Arxiv}, we model each human body as a cylinder with a radius of $R_B$ and a height of $h_B$. The bottom center of these cylinders is modeled by a 2D-PPP with the density $\lambda_B$. As illustrated in \cite[Fig. 3]{Tang2024Arxiv}, if the bottom center of a human body lies within the line-of-sight (LoS) blockage zone, the AP-UE signal transmission link is blocked. Conversely, if a human body is entirely outside the LoS blockage zone, the AP-UE link is considered as a LoS link, and the signal transmitted by the AP reaches the UE.

We employ the tractable Boolean scheme of straight lines to model wall blockages in the indoor environment, as in \cite{Shafie2021JSAC}. We assume that the length of walls, $L_W$, follow an arbitrary probability density function (PDF) of $f_{L_W}(L_W)$ with the mean $\E[L_W]$, and the centers of walls form a PPP with a density of $\lambda_W$. We ignore the widths of the walls, as their widths are much smaller compared to their lengths. Also, we assume that the orientations of the walls are binary, taking values of either $0$ or $\pi/2$ with equal probability, ensuring that the walls are either parallel or orthogonal to each other. Finally, we assume that the height of walls is fixed and equals the height of the APs, i.e., $h_W = h_A$. 

According to \cite{Wu2021TWC,Shafie2021JSAC}, the probability that an AP-UE link is not blocked by human bodies and wall blockages is given by $p_{H,\mathrm{LoS}}(d) =\exp(-\alpha d)$ and $p_{W,\mathrm{LoS}}(d) = \exp(-\eta d)$, respectively, where $\alpha \delequal 2\lambda_B R_B (h_B-h_U)/(h_A-h_U)$, $\eta = \frac{2\lambda_W\E[L_W]}{\pi}$, and $d$ is the horizontal distance between the AP and the UE. Thus, the probability that an AP-UE link is not blocked is given by $p_{\mathrm{LoS}}(d) = p_{H,\mathrm{LoS}}p_{W,\mathrm{LoS}} = \exp(-(\alpha+\eta)d)$.

In the considered system, we adopt the nearest-LoS-AP association strategy, i.e., each UE associates with its nearest AP that has a LoS propagation path to it. For the typical UE, we denote its associated AP by $AP_0$ and the horizontal distance between $U_0$ and $AP_0$ by $d_0$. For other LoS non-associated APs, we denote $AP_{i}$ as the $i$th nearest LoS non-associated AP of $U_0$ and $d_i$ as the distance between $U_0$ and $AP_i$, where $\Psi_{AP} = \{AP_1,AP_2,\cdots\}$.

\vspace{-0.1em}

\subsection{Antenna Model}

\begin{figure}[t]
    \centering
        \subfigure[The top view.]{
            \includegraphics[width=0.38\columnwidth]{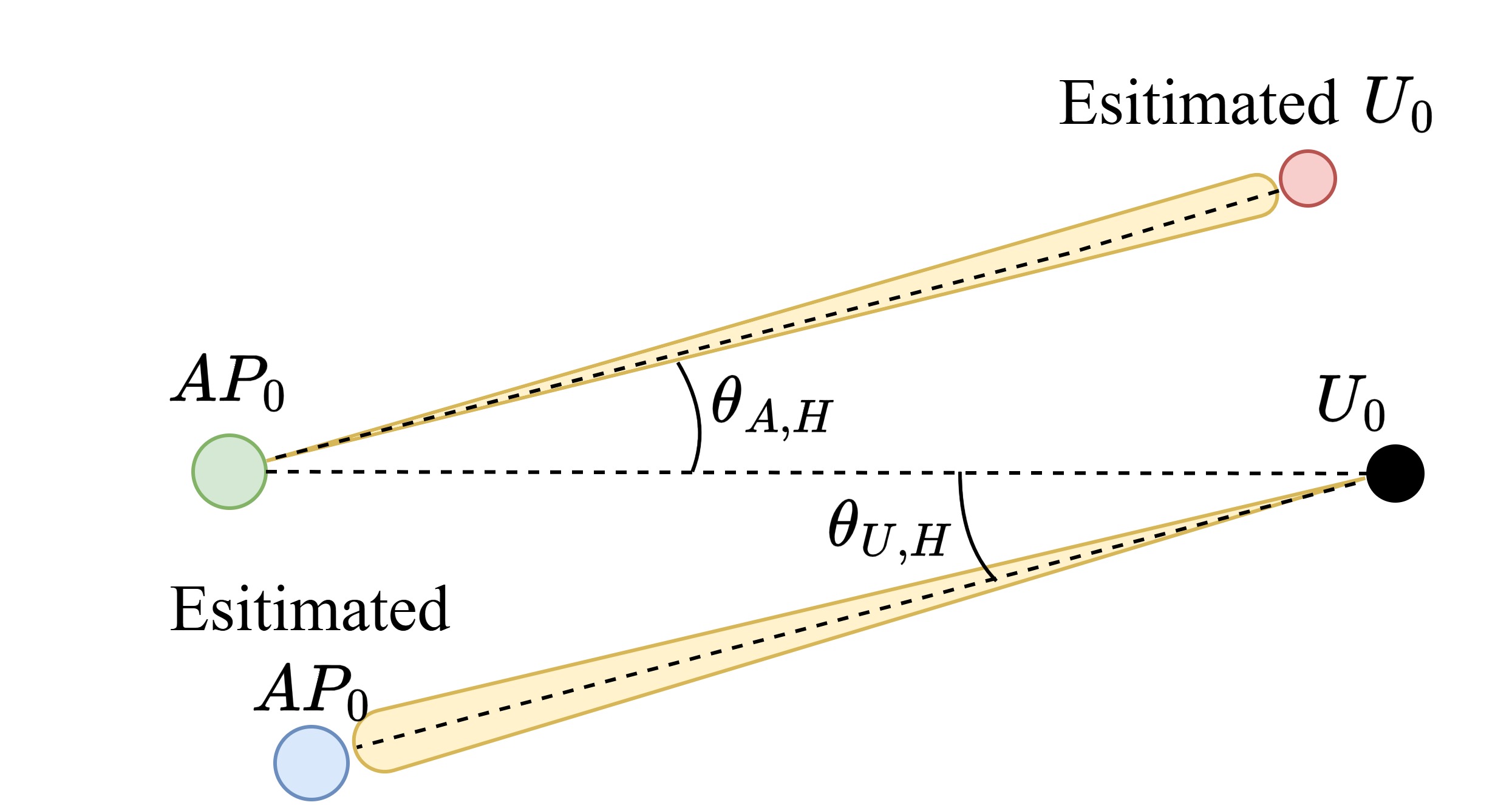}
            }\label{fig:TopviewB}
        \subfigure[The vertical view.]{
            \includegraphics[width=0.38\columnwidth]{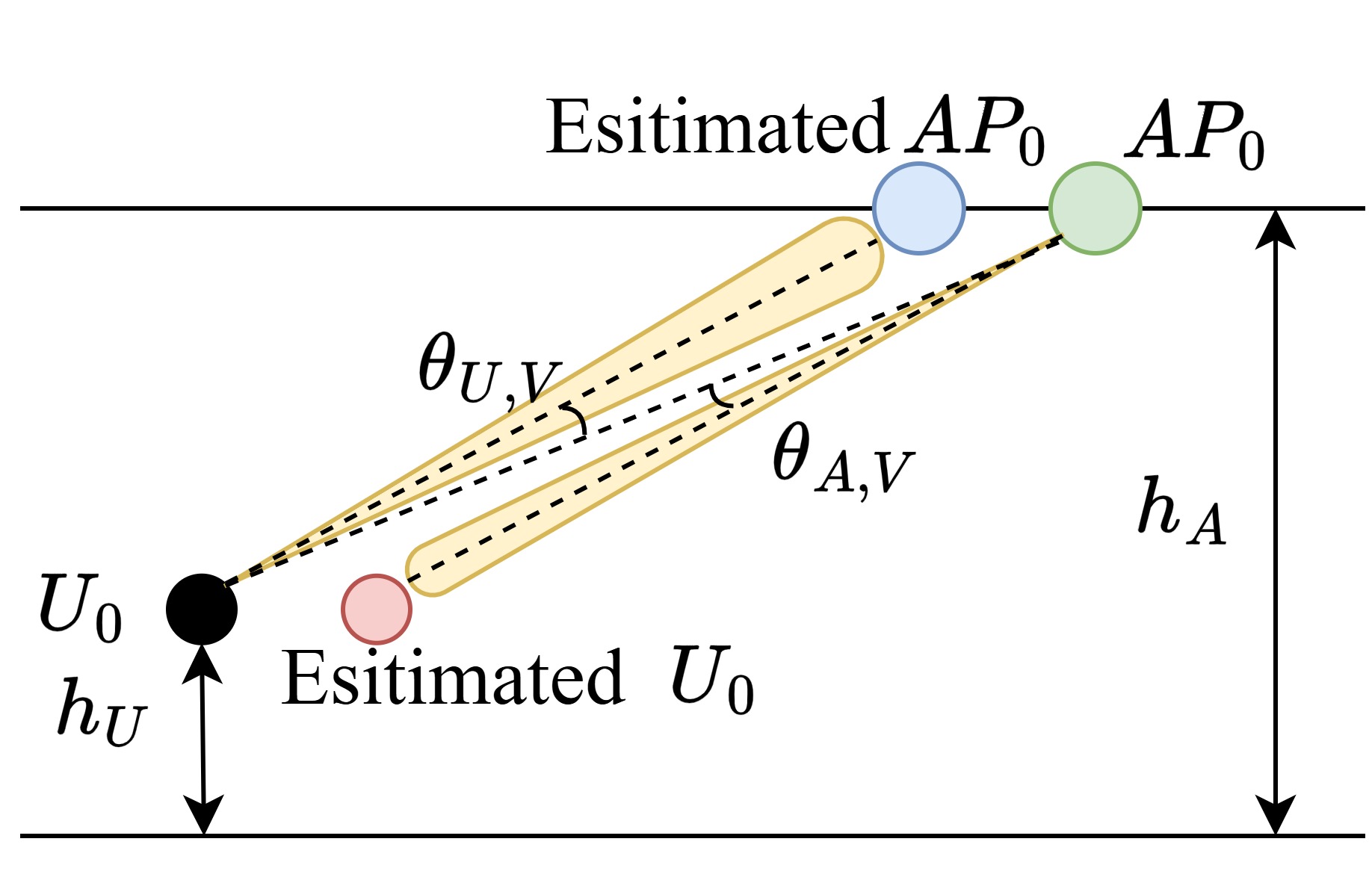}
            }\label{fig:VertviewB}
            \vspace{-0.5em}
        \caption{Top and vertical views of the pointing error for an AP-UE link.}\vspace{-2em}
        \label{fig:Beam}
\end{figure}

We assume that both APs and UEs are equipped with planar antennas to enhance signal strength to compensate for the severe path loss in THz propagation \cite{Yuan2020}. Specifically, each AP is equipped with an $N_A\times N_A$ planar antenna array, while each UE is equipped with an $N_U\times N_U$ planar antenna array, both with an element spacing of $z=\frac{\lambda}{2}$, to achieve narrow beamwidth and high antenna gain. According to \cite{Dabiri2022wcl}, the antenna gain in the direction of the elevation angle $\theta_q$ and azimuth angle $\phi_q$ is given by $G_{q}(\theta_q,\phi_q)=G_{q,\max}H_{pe,q}(\theta_q,\phi_q)$, where $G_{q,\max}\approx \pi N_q^2$ represents the maximum antenna gain, $\theta_q\!=\!\arctan\left(\!\sqrt{\tan^2(\theta_{q,V})\!+\!\tan^2(\theta_{q,H})}\!\right)$, with $\theta_{q,V}$ and $\theta_{q,H}$ representing the angles from the target direction in the horizontal and vertical plane, respectively, as shown in Fig.~\ref{fig:Beam}, and
\begin{align}
    &H_{pe,q}(\theta_q,\phi_q) \notag\\
    &\!\!=\!\left(\!\frac{\sin\left(\frac{N_{q}\pi\sin(\theta_q)\sin(\phi_q)}{2}\right)\sin\left(\frac{N_{q}\!\pi\sin(\theta_q)\cos(\phi_q)}{2}\right)}{N_{q}\sin\left(\frac{\pi\sin(\theta_q)\!\sin(\phi_q)}{2}\right)\!N_{q}\!\sin\left(\frac{\pi\sin(\theta_q)\!\cos(\phi_q)}{2}\right)}\!\right)^2.
\end{align}
Here, $q\in\{A_i,U_i\}$, where $A_i$ corresponds to $AP_i$ for the $AP_i$-$U_0$ link and $U_i$ corresponds to $U_0$ for the $AP_i$-$U_0$ link.

To enable signal transmission, a typical UE, $U_0$, initiates this process by sending a request to its associated AP, $AP_0$. Upon receiving this request, $AP_0$ estimates the location of $U_0$ and adjusts its antenna beam towards the estimated location of $U_0$. Simultaneously, $AP_0$ transmits its relative location back to $U_0$, enabling $U_0$ to adjust its antenna beam towards $AP_0$ for efficient signal reception.

\subsection{THz Channel Model}

The signal propagation at THz frequencies is affected by the distance-dependent large-scale fading and multipath-induced small-scale fading. By combining large-scale and small-scale fading models, the received power at $U_0$ from $AP_i$ is expressed as
\begin{align}
    P_i =   P_t G_{i} H_{L,i}H_{S,i},
\end{align}
where $P_t$ is the transmit power, $G_i=G_{i,A}G_{i,U}$ is the combined effective antenna gain, with $G_{i,A}$ and $G_{i,U}$ representing the antenna gains at $AP_i$ and $U_0$, respectively, and $H_{L,i}$ and $H_{S,i}$ denote large-scale and small-scale fading gains for the $AP_{i}$-$U_{0}$ link, respectively.

In THz communication systems, the large-scale fading is primarily determined by the spreading loss and the molecular absorption loss. The large-scale fading gain, $H_{L,i}$, from $AP_i$ located at the distance $d_i$ in the 3D indoor environment is expressed as $H_{L,i} = \xi W(d_{i})$, where $\xi\delequal \frac{c^2}{(4\pi f)^2}$, $c=3\times10^8$ m/s is the light speed, $f$ is the operating frequency, $W(d_{i})=\frac{1}{d_{i}^2+(h_A-h_U)^2}\exp\left(-\epsilon(f)\sqrt{d_{i}^2+(h_A-h_U)^2}\right)$, and $\epsilon(f)$ is the molecular absorption coefficient of frequency $f$. %

For the small-scale fading, compared to other existing distributions, the multi-cluster FTR (MFTR) distribution is particularly well-suited for THz communications due to the dominance of LoS and single reflected paths accompanied by diffraction and scattering at THz frequencies for short-distance transmission \cite{Papasotiriou2021}. Therefore, we model $H_{S,i}$ using the MFTR distribution. Under this model, the cumulative distribution function (CDF) of $H_{S,i}$ is given by
%\begin{align}\label{eq:pdfFTR}
%    f_{H_{S,i}}(h) \!=\!  \frac{m^m}{\Gamma(m)}\sum\limits_{j=0}^{\infty}\frac{(\mu K)^j r_j}{j!\Gamma(j+\mu)(2\sigma^2)^{j+\mu}}h^{j+\mu-1}e^{-\frac{h}{2\sigma^2}}
%\end{align}
%and
\begin{align}\label{eq:cdfFTR}
     F_{H_{S,i}}(h) =1-\frac{m^m}{\Gamma(m)}\sum\limits_{j=0}^{\infty}\frac{(\mu K)^j r_j}{j!\Gamma(j+\mu)}\Gamma\left(j+\mu,\frac{h}{2\sigma^2}\right),
\end{align}
%respectively \cite{Sanchez2024TWC}, 
where $\Gamma(\cdot)$ is the gamma function and $\Gamma(\cdot,\cdot)$ is the upper incomplete gamma function \cite{Sanchez2024TWC}. In \eqref{eq:cdfFTR}, the parameter $m$ characterizes the severity of fading, $\mu$ is the number of cluster, $K$ is the average power ratio of the dominant component to the remaining diffuse multipath, $2\sigma^2=\frac{1}{\mu (K+1)}$ is the average power of the diffuse component over MFTR fading, and $r_j$ is given in \cite[Eq.(7)]{Tang2024Arxiv},%by
%\begin{align}
 %   r_j \!&=\!\frac{(1-\Delta)^j\Gamma(m+j)}{\sqrt{\pi}(\mu K(1\!-\!\Delta)\!+\!m)^{m+j}}\sum\limits_{k=0}^j\! {\binom{j}{k}}\frac{\Gamma(k+\frac{1}{2})}{\Gamma(k+1)}\left(\frac{2\Delta}{1-\Delta}\right)^{k}\notag\\
%   & \times _2\!F_1\left(m+j,k+\frac{1}{2};k+1;\frac{-2\mu K \Delta}{\mu K(1-\Delta)+m}\right),
%\end{align}
where $\Delta$ is the parameter representing dominant waves similarity in cluster 1 of MFTR and $_2F_1(\cdot,\cdot;\cdot;\cdot)$ is the Gauss hypergeometric function.

To analyze the downlink coverage performance of a THz communication system, we employ the coverage probability, denoted by $P_c$, as our performance metric. It is defined as the probability that the received signal-to-interference-plus-noise ratio (SINR) at $U_0$ exceeds a given threshold $\gamma_{\mathrm{th}}$, i.e., $P_c = \mathrm{Pr}(\mathrm{SINR}>\gamma_{\mathrm{th}})$. Accordingly, the SINR of $U_0$ is formulated as
\begin{align}\label{eq:SINRequation}
    \mathrm{SINR} = \frac{P_{0}}{I + N_0}%=\frac{P_{0}}{\sum\limits_{AP_i\in \Psi_{AP}} P_{i} + N_0}\notag\\
    = \frac{P_t\xi G_{0}W(d_{0})H_{S,0}}{\sum\limits_{AP_i\in \Psi_{AP}} P_t\xi G_{i}W(d_{i})H_{S,i} + N_0},
\end{align}
where $I$ represents the interference power from all LoS non-associated APs and $N_0$ denotes the noise power.

\section{Coverage Probability Analysis}\label{Sec:Coverage}

In this section, we analyze the coverage probability, $P_c$, under the impact of the pointing error. In our considered system, the pointing error degrades the antenna gain for the $AP_0$-$U_0$ link, which consequently affects the received signal power. Therefore, we commence our analysis by evaluating the impact of the pointing error on the received signal power.

\subsection{Pointing Error Analysis}

%In this subsection, we first analyze $f_{G_0}(g_0)$.

The pointing error refers to the antenna direction mismatch between the transmitter and receiver. We note that the antenna beams of $U_0$ and $AP_0$ are supposed to be directed towards each other, i.e., $\theta_{A_0} = \theta_{U_0} =0$, and the antenna gain of the received signal at $U_0$ is supposed to be equal to the maximum antenna gain. However, due to the location estimation inaccuracy induced by pointing error, the antenna gain of the received signal at $U_0$ is expressed as
\begin{align}\label{eq:G0original}
    G_{0} = G_{\max}H_{pe} = G_{A,\max}G_{U,\max}H_{pe},
\end{align}
where $H_{pe}=H_{pe,A_0}(\hat\theta_{A_0},\phi_{A_0})H_{pe,U_0}(\hat\theta_{U_0},\phi_{U_0})$ represents the pointing error loss, and $\hat\theta_{A_0}$ and $\hat\theta_{U_0}$ are estimated angle error for $AP_0$ and $U_0$, respectively. Since the estimated angle error for $AP_0$ and $U_0$ originates from the location estimation of $U_0$, we assume that both $AP_0$ and $U_0$ share the same estimated angle errors, i.e., $\hat\theta_{U_0} = \hat\theta_{A_0} = \hat\theta$. Also, since square antenna arrays are deployed at the AP and the UE, we assume that the array element has a symmetrical radiation pattern in the azimuth direction \cite{Dabiri2022wcl}. According to \cite{Gerstoft2016spl}, we assume that the estimated angle error in the horizontal and vertical plane follows an independent and identical zero-mean Gaussian distribution with standard deviation $\varsigma_{\theta}$, i.e., $\hat\theta_{H}\sim\mathcal{N}(0,\varsigma_{\theta})$ and $\hat\theta_{V}\sim\mathcal{N}(0,\varsigma_{\theta})$. Given these assumptions, we derive the PDF of the pointing error loss in the following lemma.

\begin{Lemma}\label{Lemma:PointingError}
    The PDF of pointing error loss, $f_{H_{pe}}(h_{pe})$, is approximated as 
    \begin{align}\label{eq:fhpe}
        f_{H_{pe}}(h_{pe}) \approx \beta h_{pe}^{\beta-1},
    \end{align}
    where $\beta = \frac{1.06^2}{2\varsigma_{\theta}^2(N_A^2+N_U^2)}$.
    %respectively.
    \begin{IEEEproof}
    The main lobe gain at $AP_0$ and $U_0$ can be approximated by a Gaussian beam as $ H_{pe,q}(\hat\theta,\phi_{q})\approx \exp\left(-\hat\theta^2/\omega_q^2\right)$,
     %\begin{align}
    %    H_{pe,q}(\hat\theta,\phi_{q})\approx \exp\left(-\frac{\hat\theta^2}{\omega_q^2}\right),
    %\end{align}
    where $\omega_q = \frac{1.06}{N_q}$. Therefore, $H_{pe}$ is approximated by
    \begin{align}\label{eq:HpeApp}
        H_{pe}\approx \exp\left(-\frac{\hat\theta^2(N_A^2+N_U^2)}{1.06^2}\right).
    \end{align}   
    For small values of $\hat\theta_{V}$ and $\hat\theta_{H}$, $\hat\theta$ can be approximated by $\hat\theta = \sqrt{\hat\theta_{V}^2+\hat\theta_{H}^2}$. Since $\hat\theta_{V}$ and $\hat\theta_{H}$ follow independent and identical Gaussian distributions, $\hat\theta$ follows a Rayleigh distribution. Accordingly, $\hat\theta^2$ follows an exponential distribution and its PDF is given by
    \begin{align}\label{eq:PDFTheta2}
        f_{\hat\theta^2}(\Theta) = \frac{1}{2\varsigma_{\theta}^2}\exp\left(-\frac{\Theta}{2\varsigma_{\theta}^2}\right).
    \end{align}
    By combining \eqref{eq:HpeApp} and \eqref{eq:PDFTheta2}, the CDF of $H_{pe}$ is obtained as
    \begin{align}
        F_{H_{pe}}(h_{pe}) &= 1-F_{\hat\theta^2}\left(-\frac{1.06^2\ln h_{pe}}{N_A^2+N_U^2}\right)= h_{pe}^{\beta}.
    \end{align}
     We then obtain $f_{H_{pe}}(h_{pe})$ by taking the derivative of $F_{H_{pe}}(h_{pe})$ with respect to $h_{pe}$, resulting in \eqref{eq:fhpe}.
    \end{IEEEproof}
\end{Lemma}

\textit{Remark 1:} From Lemma~\ref{Lemma:PointingError}, we find that the average antenna gain under the asymptotic condition is given by
\begin{align}
    \lim\limits_{N_A\!\rightarrow\!\infty}\!\overline{G}_{0} \!&=\! \lim\limits_{N_A\!\rightarrow\!\infty}\frac{1.06^2\pi^2N_A^2N_U^2}{1.06^2\!+\!2(N_A^2\!+\!N_U^2)\varsigma_{\theta}^2}\! =\! \frac{1.06^2\pi^2N_U^2}{2\varsigma_{\theta}^2},
\end{align}
which is constrained by $\varsigma_{\theta}$ and extremely lower than the maximum antenna gain of AP, which approaches infinity. It indicates that the pointing error has a significant impact on the received signal power, and in turn, impacts the coverage probability. Additionally, the expression for $\overline{G}_{0}$ reveals that the average antenna gain decreases when the difference between $N_A$ and $N_U$ increases while $N_A N_U$ is kept as a constant.

%We note that the pointing error is affected by the uplink sensing, which . The improve of the transmission antenna may lead to limited improve of the coverage performance. 
\vspace{-1em}
\subsection{AP Association and Interference Analysis}

In this subsection, we derive the PDF of the horizontal distance from $U_0$ to $AP_0$, $f_{D_0}(d_0)$, and evaluate the antenna gain of interference. Leveraging the properties of the PPP, the AP intensity at a horizontal distance $d$ from $U_0$ is given by $\Lambda_{A}(d) = 2\pi d\lambda_{A}$. Based on this, we first derive $f_{D_0}(d_0)$ in the following Lemma.
\begin{Lemma}\label{Lemma:intensity}
    The PDF of the horizontal distance between $U_0$ and $AP_0$, $d_0$, is derived as
    \begin{align}\label{eq:pdfLoSAP}
        f_{D_0}(d_0) = \Lambda_{A,\mathrm{LoS}}(d_0) e^{-\int_{0}^{d_0}\Lambda_{A,\mathrm{LoS}}(d) \mathrm{d}d},
    \end{align}
    where $\Lambda_{A,\mathrm{LoS}}(d_0) = \Lambda_{A}(d_0)e^{-(\alpha+\eta) d_0 }$.
    \begin{IEEEproof}
    Based on the blockage model, the LoS AP intensity at a horizontal distance $d$ from $U_0$ is given by $\Lambda_{A,\mathrm{LoS}}(d) = e^{-\alpha d }\Lambda_{A}(d)$. Since APs are distributed according to a PPP, the probability that the number of LoS APs within the distance $d$ from $U_0$, denoted by $N_{A,\mathrm{LoS}}(d)$, equals $n$ is given by
\begin{align}
    \mathrm{Pr}\left(N_{A,\mathrm{LoS}}(d)=n\right) = \frac{(\Xi(d))^n}{n!}\exp(-\Xi(d)),
\end{align}
where $\Xi(d)\!=\!\int_{0}^{d}\Lambda_{A,\mathrm{LoS}}(x) \mathrm{d}x$. Thus, the CDF of $d_0$ is obtained as $ F_{D_0}(d_0)\!=\! 1\!-\!\exp(\!-\Xi(d_0)\!)$. By taking the derivative of $F_{D_0}(d_0)$, we obtain $f_{D_0}(d_0)$ as given in \eqref{eq:pdfLoSAP}.
%, which is given by $f_{D_0}(d_0) = \frac{\mathrm{d}F_{D_0}(d_0)}{\mathrm{d}d_0}$, resulting in the expression in \eqref{eq:pdfLoSAP}. 
\end{IEEEproof}
\end{Lemma}

Based on Lemma~\ref{Lemma:intensity}, we find that the average horizontal distance between $U_0$ and $AP_0$, $d_0$, increases monotonically with both the average wall length and the density of human blockages. This indicates that greater obstruction in the environment pushes the associated AP farther from the user, due to the reduced LoS probability and increased attenuation in nearby areas.

We then analyze antenna gains of interference signals, laying the foundations for our subsequent interference analysis. The antenna gain of the transmit signal from $AP_i$ to $U_0$ is the product of transmit and receive antenna gains, expressed as $G_i=G_{i,A}G_{i,U}$. To simplify the antenna gain analysis of interfering APs, we adopt the cone model proposed in \cite{Shafie2021JSAC}. We assume that the vertical and horizontal main lobe beamwidths of APs and UEs are equal to their half-power beamwidth, i.e., $\phi_{q,V} = \phi_{q,H}\approx \frac{0.886\times2}{N_q}$. Moreover, the antenna gains of the main lobe and the side lobes for APs and UEs are expressed as $G_{q}^m = \pi N_q^2$ and
\begin{align}
    G_{q}^s = \frac{\pi-N_q^2\pi\arcsin\left(\tan\left(\frac{\phi_{q,V}}{2}\right)\tan\left(\frac{\phi_{q,H}}{2}\right)\right)}{\pi-\arcsin\left(\tan\left(\frac{\phi_{q,V}}{2}\right)\tan\left(\frac{\phi_{q,H}}{2}\right)\right)},
\end{align}
respectively. Based on these assumptions, we examine transmit and receive antenna gains separately.

\subsubsection{Transmit Antenna Gain}
The transmit antenna gain depends on whether $U_0$ falls within the antenna beam of the interfering AP. We assume that the depression angle from the AP is uniformly distributed over $[\phi_{AP},\pi/2]$, where $\phi_{AP} = \arctan\left(\frac{h_A-h_U}{R_A}\right)$ is the depression angle from the AP to its coverage boundary with radius $R_A$ \cite{Wu2021TWC}. We also assume that the horizontal beam direction from the AP is uniformly distributed over $[0,2\pi)$. Therefore, the transmit antenna gain equals the main lobe gain with the hitting probability $p_{A} = p_{A,V}p_{A,H}$, where $p_{A,V}=\min\left\{\frac{\phi_{A,V}}{\frac{\pi}{2}-\phi_{AP}},1\right\}$ and $p_{A,H} = \frac{\phi_{A,H}}{2\pi}$ are the probabilities that $U_0$ is located within the AP's vertical beam and horizontal beam, respectively.

\subsubsection{Receive Antenna Gain}
The receive antenna gain depends on whether the interfering AP is within the antenna beam of $U_0$. We note that the antenna beam of $U_0$ is oriented towards its associated AP, $AP_0$. Similar to the analysis of the transmit antenna gain, the receive antenna gain is equal to the main lobe gain with the hitting probability $p_U = p_{U,V}p_{U,H}$, where $p_{U,V}$ and $p_{U,H}$ are the probabilities that the interfering AP is within the vertical and horizontal beams of $U_0$, respectively. As shown in \cite[Fig. 6(a)]{Tang2024Arxiv}, the vertical beam range of $U_0$ is determined by the maximum horizontal distance $R_{U_0,\text{max}}$, which is calculated as
\begin{align}
    R_{U_0,\text{max}} = \frac{h_A-h_U}{\tan\left(\arctan\left(\frac{h_A-h_U}{d_0}\right)-\frac{\phi_{U,V}}{2}\right)}.
\end{align}
An interfering AP, $AP_i$, is within the vertical beam of $U_0$ if its distance to $U_0$, $d_i$, is less than or equal to $R_{U_0,\text{max}}$. For the horizontal beam of $U_0$, based on the properties of the PPP, we have $p_{U,H} = \frac{\phi_{U,H}}{2\pi}$. By combining $p_{U,V}$ and $p_{U,H}$, we obtain the hitting probability of an interfering AP, $AP_i$, within the antenna beam of $U_0$.

Based on our analysis of both transmit and receive antenna gains, the probability distribution of the antenna gain for $AP_i$ is given by
\begin{align}\label{eq:PRGi}
    \mathrm{Pr}(G_i) =\left\{
    \begin{aligned}
        &p_Ap_U, &&\text{if }G_i=G_{A,m} G_{U,m},\\
        &p_A(1-p_U), &&\text{if }G_i=G_{A,m} G_{U,s},\\
        &(1-p_A)p_U, &&\text{if }G_i=G_{A,s} G_{U,m},\\
        &(1-p_A)(1-p_U), &&\text{if }G_i=G_{A,s} G_{U,s}.
    \end{aligned}
    \right.
\end{align}

\subsection{Coverage Performance Analysis}

Building on the derived $f_{H_{pe}}(h_{pe})$, $f_{D_0}(d_0)$, and antenna gain analysis for interference signals, we now derive the coverage probability and present it in the following Theorem.

\begin{Theorem}\label{Theorem:Coverage}
   The coverage probability of $U_0$ is derived as
    \begin{align}\label{eq:PcTheorem}
       & P_{c} = \int_0^{\infty} \int_0^1 f_{H_{pe}}(h_{pe}){\Gamma(m)}\sum\limits_{j=0}^{\infty}\frac{(\mu K)^j r_j}{\Gamma(j+1)}\sum\limits_{l=0}^{j+\mu-1}\frac{(-s)^l}{l!}\notag\\
        &\times\frac{\partial^{(l)}\Lp_{I+N_0}(s|h_{pe},d_0)}{\partial s^l}\Bigg|_{s=\rho} f_{D_0}(d_0)\mathrm{d}h_{pe}\mathrm{d}d_0,
    \end{align}
where $\rho = \frac{(j+\mu)\gamma_{\mathrm{th}}}{2\sigma^2 P_t\xi G_{\max}h_{pe} W(d)}$ and 
\begin{align}\label{eq:LaplaceTheorem}
    &\Lp_{I+N_0}(s|h_{pe},d_0)\! =\!\exp\Bigg(\!\!-\!sN_0\!-\!\int_{d_0}^{\infty}\Lambda_{A,\mathrm{LoS}}(d)\sum\limits_{G_i}\mathrm{Pr}(G_i)\notag\\
    &\times\Bigg(\frac{m^m}{\Gamma(m)}\sum\limits_{j=0}^{\infty}\frac{(\mu K)^j r_j }{j!\left(1+2s\sigma^2 P_t \xi G_{i} W(d)\right )^{j+\mu}}\Bigg)\mathrm{d}d\Bigg)
\end{align}
represents the Laplace transform of the interference plus noise power.%, $\mathrm{Pr}(G_i)$ is given in \eqref{eq:PRGi}, and $f(d_0)$ is given in \eqref{eq:pdfLoSAP}.
\begin{IEEEproof}
We note that the coverage probability can be calculated by
\begin{align}\label{eq:Pc}
    P_c = &\int_{0}^{\infty}\int_{0}^{1}\mathrm{Pr}\left(\frac{P_t \xi G_{\max}h_{pe}W(d_{0})H_{S,0}}{I + N_0}>\gamma_{\mathrm{th}}\bigg|h_{pe},d_0\right)\notag\\
    &\times f_{H_{pe}}(h_{pe})f_{D_0}(d_0)\mathrm{d}h_{pe}\mathrm{d}d_0.
\end{align}
Using the CDF of $H_{S,0}$ in \eqref{eq:cdfFTR}, the conditional coverage probability in \eqref{eq:Pc} is expressed as
\begin{align}\label{eq:PrH0geqT1}
    &\mathrm{Pr}\left(\frac{P_t \xi G_{\max}h_{pe}W(d_{0})H_{S,0}}{I + N_0}>\gamma_{\mathrm{th}}\bigg|h_{pe},d_0\right)\notag\\
    %&= \E_I\left[\mathrm{Pr}\left(H_0>\frac{\beta}{g_0 W(d_0)}(I +N)\Bigg|d_0,I\right)\right]\notag\\
    &=\E_I\left[\frac{m^m}{\Gamma(m)}\!\sum\limits_{j=0}^{\infty}\frac{(\mu K)^j r_j}{\Gamma(j\!+\!1)}\!\sum\limits_{l=0}^{j\!+\!\mu\!-\!1}\frac{\left(\rho(I+N_0)\right)^l}{l!}e^{-\rho(I+N_0)}\right]\notag\\
    &= \!\frac{m^m}{\Gamma(\!m\!)}\!\!\sum\limits_{j=0}^{\infty}\!\frac{(\!\mu K\!)^j r_j}{\Gamma(\!j\!+\!1\!)}\!\sum\limits_{l=0}^{j\!+\!\mu\!-\!1}\frac{(\!-\!s\!)^l}{l!}\!\frac{\partial^{(l)}\!\Lp_{I\!+\!N_0}(\!s|h_{pe},\!d_0\!)}{\partial s^l}\Bigg|_{s\!=\rho}.
\end{align}
According to the property of the PPP, we calculate $\Lp_{I\!+\!N_0}(s|h_{pe},d_0)$ in \eqref{eq:PrH0geqT1} as
\begin{align}\label{eq:LaplaceIN}
    &\Lp_{I\!+\!N_0}(s|h_{pe},d_0)=e^{-sN_0}\E\left[\exp\left(-s\sum\limits_{AP_i\in\Psi_{AP}}P_i\right)\Bigg|d_0\right]\notag\\
    &\!=\!\exp\left(\!-\!sN_0\!-\!\!\int_{d_0}^{\infty}\!(1\!-\!\E\left[\exp\left(\!-\!s P_i\right)\right])\Lambda_{A,\mathrm{LoS}}(d)\mathrm{d}d\right),
\end{align}
where $\E\left[\exp\left(-s P_i\right)\right]$ is computed as
\begin{align}\label{eq:Espidid}
    &\E\left[\exp\left(-s P_i\right)\right]=\sum\limits_{G_i}\mathrm{Pr}(G_i)\notag\\
    &\times\Bigg(\frac{m^m}{\Gamma(m)}\sum\limits_{j=0}^{\infty}\frac{(\mu K)^j r_j }{j!\left(1+2s\sigma^2 P_t \xi G_{i} W(d)\right )^{j+\mu}}\Bigg).
\end{align}
By substituting \eqref{eq:LaplaceTheorem} into \eqref{eq:PrH0geqT1} and combining it with \eqref{eq:Pc}, we obtain the coverage probability of $U_0$ as \eqref{eq:PcTheorem}.
\end{IEEEproof}
\end{Theorem}

From Theorem~\ref{Theorem:Coverage}, we find that the pointing error has a significant impact on the coverage performance. This impact can be addressed by employing technologies that enhance angle-of-arrival detection, enabling more accurate estimation of the UE’s location. This improved estimation leads to more precise antenna alignment, which in turn increases the received signal power and ultimately enhances the coverage probability.
%This improvement in angle estimation leads to better alignment between the antennas, which in turn increases the receiver's signal power, thereby improving the coverage probability.

\section{Numerical Results}\label{Sec:Num}

In this section, we first present numerical results to verify our analysis in Section~\ref{Sec:Coverage} and evaluate the impact of various parameters on the coverage probability, such as the standard deviation of the estimated angle error and the number of antennas at the AP and UE. The values of the parameters used in this section are summarized in Table~\ref{tab:System_Para}, unless specified otherwise.

\begin{table} 
\centering
\caption{Value of System Parameters Used in Section~\ref{Sec:Num}.}
\vspace{-0.5em}
\begin{tabular}{|l|l|l|}
\hline
    \textbf{Parameter} & \textbf{Symbol} & \textbf{Value} \\
    \hline
  Operating frequency  &  $f$ & $0.3$ THz \\\hline
    Absorption coefficient &  $\epsilon(f)$ & $0.00143$ $\mathrm{m}^{-1}$ \\\hline
   Height of APs and UEs  &  $h_A$, $h_U$ & $3$ m, $1$ m \\\hline
   Height and radius of human &$h_B$, $R_B$ & $1.7$ $\mathrm{m}$, $0.25$ $\mathrm{m}$\\\hline
    Density of APs and human  &  $\lambda_A$, $\lambda_B$ & $0.1$ $\mathrm{m}^{-2}$, $0.1$ $\mathrm{m}^{-2}$ \\\hline
     Density and length of wall &  $\lambda_W$, $\E[L_W]$ & $0.04$ $\mathrm{m}^{-2}$, $3$ $\mathrm{m}$ \\\hline
    Transmit and AWGN power & $P_t$, $N_0$ & $5$ dBm, $-77$ dBm\\\hline
     Antenna parameters & $N_{A}$, $N_{U}$, $R_A$  & $16$, $4$, $15$ $\mathrm{m}$  \\\hline
   MFTR fading parameters & $K$, $m$, $\Delta$, $\mu$ & $5$, $2$, $0.3$, $2$ \\\hline
  
    %UE's antenna parameters & $G_U^{m}$, $G_{u}^{s}$, $k_U$, $\phi_{U,H}$, $\phi_{U,V}$ & $15$ dBi, $-10$ dBi, $0.1$, $33^{\circ}$,$33^{\circ}$ \\\hline
\end{tabular}
\vspace{-1.6em}
\label{tab:System_Para}
\end{table}

\begin{figure}[t!]
    \centering
    \includegraphics[width=0.7\columnwidth]{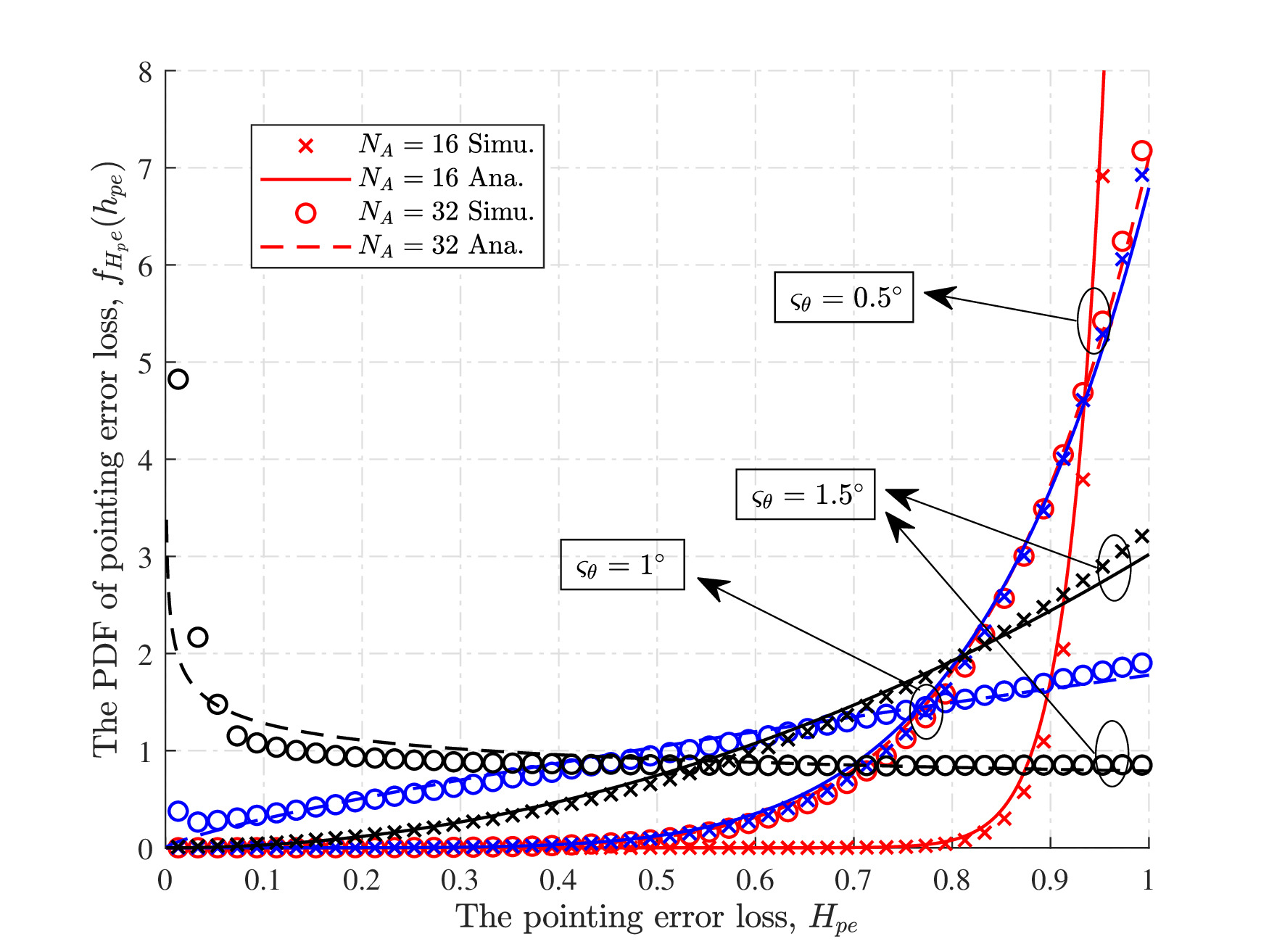}
    \vspace{-1.2em}
    \caption{The PDF of the pointing error loss $f_{H_{pe}}(h_{pe})$.}
    \label{fig:Hpe}\vspace{-2em}
\end{figure}

Fig.~\ref{fig:Hpe} plots the PDF of the pointing error loss, $f_{H_{pe}}(h_{pe})$. We first observe that our analytical results in Lemma~\ref{Lemma:PointingError} align well with the simulation results, validating the correctness of our analysis. We then observe that this PDF for $N_A=32$ is larger than that for $N_A=16$ when $H_{pe}$ is small, but smaller when $H_{pe}$ is large. This is due to the fact that a larger number of antennas results in a narrower main lobe, which increases the likelihood of a smaller pointing error loss compared to the case with fewer antennas. This observation also indicates that the average pointing error loss decreases when the number of antennas increases, highlighting the importance of accurate location estimation in THz communication systems with large antenna arrays.

\begin{figure}[t!]
    \centering
    \includegraphics[width=0.7\columnwidth]{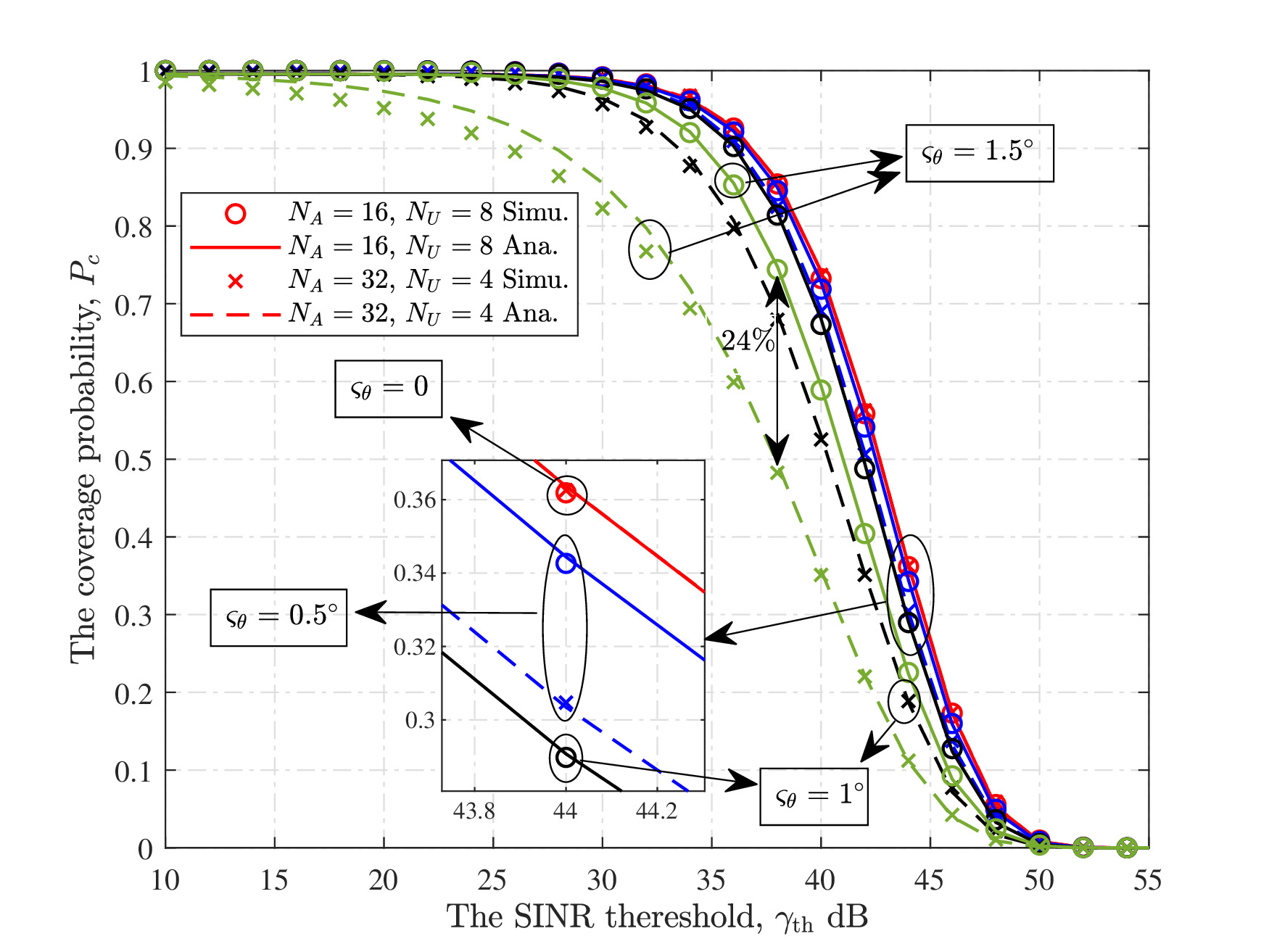}
    \vspace{-1em}
    \caption{The coverage probability, $P_c$, versus the $\mathrm{SINR}$ threshold, $\gamma_{\mathrm{th}}$ dB.}
    \label{fig:CoverP}\vspace{-1.5em}
\end{figure}

Fig.~\ref{fig:CoverP} plots the coverage probability, $P_c$, versus the $\mathrm{SINR}$ threshold, $\gamma_{\mathrm{th}}$, for various $\varsigma_{\theta}$. We first observe that our analytical results in Theorem~\ref{Theorem:Coverage} align well with the simulation results, especially for small $\varsigma_{\theta}$, confirming the accuracy of our analysis. We note that there is a small gap between the simulation and the analytical results for $\varsigma_{\theta} = 1.5^{\circ}$ at low SNR thresholds. This gap is caused by the approximation of the main lobe gain using a Gaussian beam model, which neglects side lobe gain. This small side lobe gain can noticeably affect the coverage probability at low SNR thresholds and lead to a mismatch between analytical and simulation results. We then observe that when $\varsigma_{\theta} = 0$, the coverage probability is almost identical for two antenna configurations, namely, $N_A=16$ with $N_U=8$ and $N_A=32$ with $N_U=4$. However, when $\varsigma_{\theta} = 1.5^{\circ}$, a significant difference in $P_c$ emerges between these two configurations. This observation supports the insights from Remark 1, as it reveals that while the maximum antenna gain is the same for both configurations, a larger antenna array, such as $N_A=32$, results in a narrower beam. This narrower beam makes the system more sensitive to the estimated angle error, leading to a more substantial degradation in $P_c$ as $\varsigma_{\theta}$ increases.

\begin{figure}[t!]
    \centering
    \includegraphics[width=0.7\columnwidth]{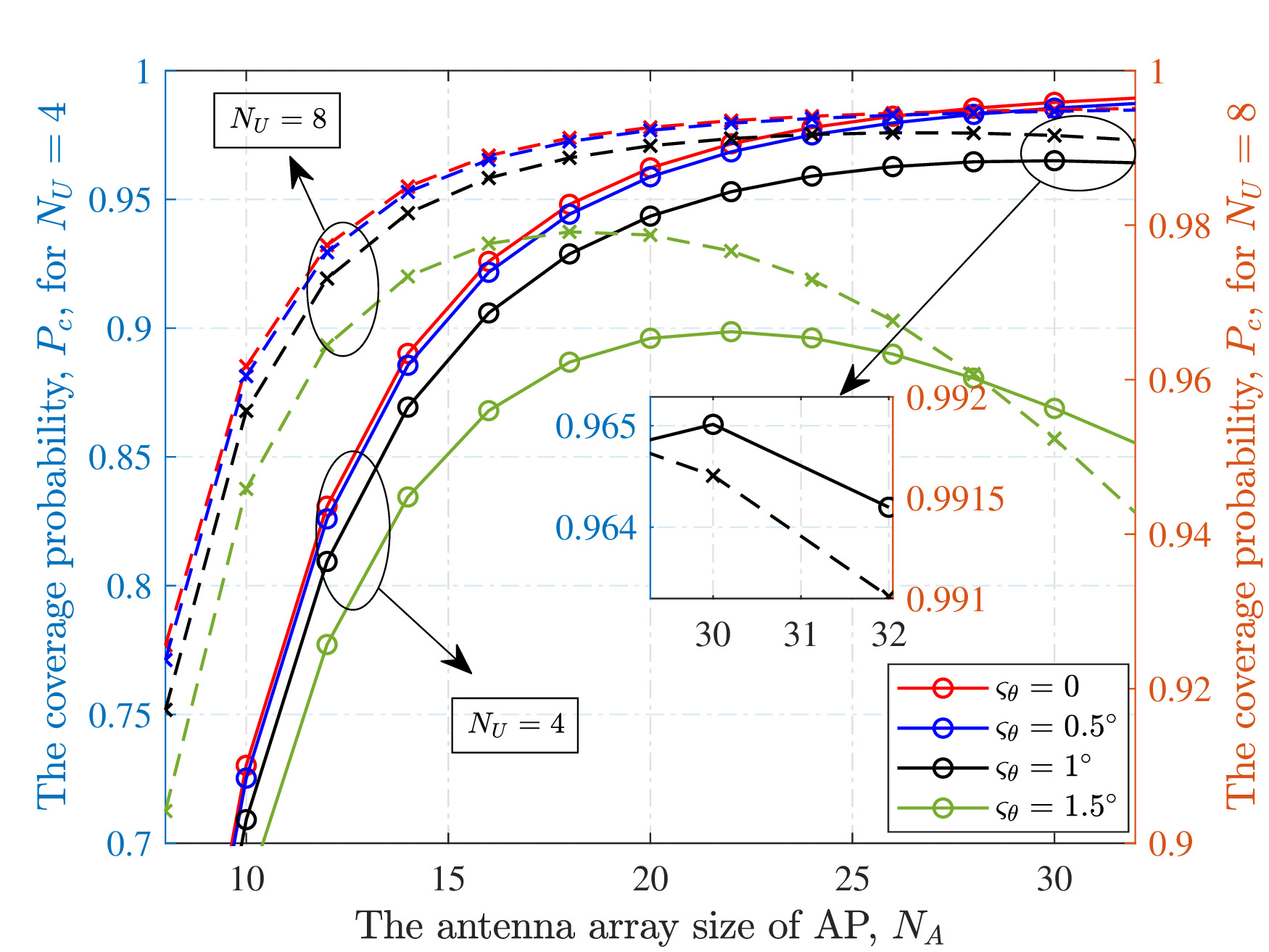}
    \vspace{-1em}
    \caption{The coverage probability, $P_c$, versus the antenna array size of AP, $N_A$, with the $\mathrm{SINR}$ threshold $\gamma_{\mathrm{th}}=30$ dB.}
    \label{fig:CoverNA}\vspace{-2em}
\end{figure}

Fig.~\ref{fig:CoverNA} plots the coverage probability, $P_c$, versus the antenna array size of AP, $N_A$, with the $\mathrm{SINR}$ threshold $\gamma_{\mathrm{th}}=30$ dB. We observe that for $\varsigma_{\theta}=0$, $P_c$ increases monotonically with $N_A$. However, when $\varsigma_{\theta}\neq0$, $P_c$ initially increases and then decreases as $N_A$ increases, particularly when $\vartheta$ is large. This observation is due to the fact that as $N_A$ increases, the antenna gain of the AP improves, enhancing the received signal power. However, as $N_A$ continues to increase, the beamwidth of the AP decreases, which in turn increases the side lobe gain, leading to a higher interference power. This effect is more pronounced as $\varsigma_{\theta}$ increases. This observation implies that the coverage probability is fundamentally constrained by the pointing error and cannot be indefinitely improved by simply increasing the antenna array size. In addition, it highlights the critical role of accurate location estimation in enhancing the coverage performance in THz communication systems.

\section{Conclusion}\label{Sec:Conclusion}

We developed a tractable analytical framework to evaluate the impact of pointing error on the coverage performance of indoor THz communication systems. We derived a new expression for the coverage probability of the typical UE, incorporating both the small-scale fading and pointing error. Through numerical results, we first verified our analysis and examined how the pointing error affects the received signal power at the UE. We then evaluated the joint impact of the pointing error and the antenna array size on the coverage probability. Our results quantified the degradation caused by the pointing error in the coverage probability, highlighting the critical importance of accurate pointing in THz systems. Furthermore, we demonstrated that merely increasing the antenna array size is insufficient to enhance coverage performance. These findings highlight the necessity of adopting accurate location estimation techniques to optimize performance in indoor THz communication systems.

\vspace{-0.2em}

\section*{Acknowledgment}
This work was funded by the Australian Research Council Discovery Project DP230100878.

%\begin{appendices}
%\section{Proof of Lemma \ref{Lemma:PointingError}}\label{Appendix:Lemma1}

%\section{Proof of Theorem \ref{Theorem:Coverage}}\label{Appendix:Theorem}

%\end{appendices}

\vspace{-0.2em}
\bibliographystyle{IEEEtran} 
\bibliography{bibli}

\end{document}